\def\bar{\overline}

\def\a{\alpha}
\def\b{\beta}

\def\e{\epsilon}

\def\bar{\overline}

\def\eV{{\rm eV}}
\def\ue3{\left| U_{e3} \right|}
\def\mnu{{\mathcal M}_{\nu f}}
\def\be{\begin{equation}}
\def\ee{\end{equation}}
\def\solm{\Delta_{\rm sun}}
\def\sola{\theta_{\rm sun}}
\def\atm{\Delta_{\rm atm}}
\def\soa{\frac{\Delta_{\rm sun}}{\Delta_{\rm atm}}}
\def\ord{{\mathcal O}}
\def\dt{\delta_\tau}
\documentstyle [a4,epsf,11pt,epsf]{article}
\begin{document}
\baselineskip=22 pt
\setcounter{page}{1}
\thispagestyle{empty}
\topskip  -2.5  cm
\begin{flushright}
\begin{tabular}{c c}
& {\normalsize  UWThPh-2004-19}\\
& {\normalsize  OCHA-PP-244}\\
&  January  19, 2005

\end{tabular}
\end{flushright}
\vspace{0.2 cm}

\centerline{\Large \bf Non-vanishing $U_{e3}$ and $\cos{2 \theta_{23}}$
from a broken $Z_2$ symmetry}
\vskip 0.8 cm
\centerline{{\large Walter Grimus,$^{a}$}$\!\!\!$
\renewcommand{\thefootnote}{\fnsymbol{footnote}}
\footnote[1]{e-mail:  walter.grimus@univie.ac.at} \
{\large Anjan S.\ Joshipura,$^{b}$}$\!\!\!$
\renewcommand{\thefootnote}{\fnsymbol{footnote}}
\footnote[2]{e-mail:  anjan@prl.ernet.in} \
{\large Satoru Kaneko,$^{c}$}$\!\!\!$
\renewcommand{\thefootnote}{\fnsymbol{footnote}}
\footnote[3]{e-mail: satoru@phys.ocha.ac.jp}}

\centerline{{\large Lu\'\i s Lavoura,$^{d}$}$\!\!\!$
\renewcommand{\thefootnote}{\fnsymbol{footnote}}
\footnote[4]{e-mail: balio@cftp.ist.utl.pt} \
{\large Hideyuki Sawanaka $^{e}$}$\!\!\!$
\renewcommand{\thefootnote}{\fnsymbol{footnote}}
\footnote[5]{e-mail: hide@muse.sc.niigata-u.ac.jp} \
{\large and  Morimitsu Tanimoto$\,^{f}$}$\!\!\!$
\renewcommand{\thefootnote}{\fnsymbol{footnote}}
\footnote[6]{e-mail: tanimoto@muse.sc.niigata-u.ac.jp}}

\vskip 0.5 cm

\centerline{$^a\!$ Institut f\"ur Theoretische Physik,
Universit\"at Wien, 
Boltzmanngasse 5,
A--1090 Wien,
Austria}

\centerline{$^b\!$ Physical Research Laboratory,
Ahmedabad 380009,
India}

\centerline{$^c\!$ Department of Physics,
Ochanomizu University,
Tokyo 112-8610,
Japan}

\centerline{$^d\!$ Instituto Superior T\'ecnico,
Universidade T\'ecnica de Lisboa,
P--1049-001 Lisboa,
Portugal}

\centerline{$^e\!$ Graduate School of  Science and Technology,
 Niigata University,  950-2181 Niigata, Japan}

\centerline{$^f\!$ Department of Physics,
Niigata University,
950-2181 Niigata,
Japan}

\vskip 0.5 cm
\centerline{\bf ABSTRACT}\par
\vskip 0.2 cm
It is shown that the neutrino mass matrices
in the flavour basis yielding a vanishing $U_{e3}$
are characterized by invariance
under a class of  $Z_2$ symmetries.
A specific $Z_2$ in this class also leads
to a maximal atmospheric mixing angle $\theta_{23}$.
The breaking of that $Z_2$ can be parameterized
by two dimensionless quantities,
$\e$ and $\e'$;
the effects of $\e, \e' \neq 0$
are studied perturbatively and numerically. 
The induced value of $\ue3$
strongly depends on the neutrino mass hierarchy.
We find that $\ue3$ is less than $0.07$ for a normal mass hierarchy,
even when $\e, \e' \sim 30 \%$.
For an inverted mass hierarchy $\ue3$ tends to be around $0.1$ but
can be as large as $0.17$.
In the case of quasi-degenerate neutrinos,
$\ue3$ could be close to its experimental upper bound $0.2$.
In contrast,
$\left| \cos{2\theta_{23}} \right|$
can always reach its experimental upper bound $0.28$.
We propose a specific model,
based on electroweak radiative corrections in the MSSM,
for $\e$ and $\e'$.
In that model, both
$\ue3$ and $\left| \cos{2 \theta_{23}} \right|$,
could be close to their respective experimental upper bounds
if neutrinos are quasi-degenerate.
\newpage

\topskip -0.5 cm
\section{Introduction}
In recent years,
the observation of solar~\cite{solar,Kamland}
and atmospheric~\cite{atmospheric} neutrino oscillations
has dramatically improved our knowledge of neutrino masses
and lepton mixing.
The neutrino mass-squared differences $\solm$ and $\atm$,
and the mixing angles $\tan^2{\sola}$
and $\sin^2{2 \theta_{\rm atm}}$,
are now quite well determined.
The third mixing angle,
represented by the matrix element $U_{e3}$
of the lepton mixing matrix $U$ (MNS matrix~\cite{mns}),
is  constrained to be small
by the non-observation of neutrino oscillations
at the CHOOZ experiment~\cite{chooz}.

In spite of all this progress,
the available information on neutrino masses and lepton mixing
is not sufficient to uncover the mechanism
of neutrino mass generation.
In particular,
we do not yet know whether the observed features of lepton mixing
are due to some underlying flavour symmetry,
or they are mere mathematical coincidences~\cite{smir}
of the seesaw mechanism.
Two features of lepton mixing
which would suggest a definite symmetry
are the small magnitudes of $U_{e3}$ and $\cos{2 \theta_{23}}$,
where $\theta_{23}$ is one of the angles
in the standard parameterization of the MNS matrix
and coincides with the atmospheric mixing angle $\theta_{\rm atm}$
when $U_{e3}=0.$
The best-fit value for $\theta_{23}$
in a two-generation analysis~\cite{atmospheric}
of the atmospheric data is $\theta_{23} = \pi / 4$,
corresponding to $\cos{2 \theta_{23}} = 0$.
Likewise,
$\ue3$ is required to be small:
$\ue3 \leq 0.26$ at $3\sigma$ from a combined analysis
of the atmospheric and CHOOZ data~\cite{lisi}.
This smallness strongly hints at some flavour symmetry.

There are many examples of symmetries
which can force $U_{e3}$ and/or $\cos{2 \theta_{23}}$ to vanish.
Both quantities vanish in the extensively studied
bi-maximal mixing \textit{Ansatz}~\cite{bd,nubim,lbim,sm},
which can be realized through a symmetry~\cite{nusmoh}. 
One can also make both $U_{e3}$ and $\cos{2 \theta_{23}}$ zero
while leaving the solar mixing angle arbitrary~\cite{mutau,d41}.
Alternatively,
it is possible to force only $U_{e3}$ to be zero,
by imposing a discrete Abelian~\cite{low}
or  non-Abelian~\cite{d42} symmetry;
conversely,
one can obtain maximal atmospheric mixing but a free $U_{e3}$
by means of a non-Abelian symmetry or a non-standard CP
symmetry~\cite{cp}. 

The symmetries mentioned above need not be exact.
It is important to consider perturbations of those symmetries
from the phenomenological point of view
and to study quantitatively~\cite{as}
the magnitudes of $U_{e3}$ and $\cos{2 \theta_{23}}$
possibly generated by such perturbations.

This paper is a study of a special class of symmetries
and of the consequences of their perturbative violation.
We show in section~2 that $U_{e3}$ vanishes
if the neutrino mass matrix in the flavour basis
is invariant under a class of $Z_2$ symmetries.
The solar and atmospheric mixing angles,
as well as the neutrino masses,
remain unconstrained by these $Z_2$ symmetries.
Those $Z_2$ symmetries thus constitute a general class of symmetries
leading only to a vanishing $U_{e3}$.
We point out that there is a special $Z_2$ in this class which leads,
furthermore,
to maximal atmospheric mixing.
We consider more closely that specific $Z_2$ in section~3,
wherein we study departures from the symmetric limit.
We parameterize perturbations of the $Z_2$-invariant mass matrix
in terms of two complex parameters,
and derive general expressions for $U_{e3}$
and $\cos{2 \theta_{23}}$ in terms of those parameters;
we also present detailed numerical estimates of $U_{e3}$
and $\cos{2 \theta_{23}}$.
Section~4 is devoted to the study of the specific perturbation
which is induced by the electroweak radiative corrections
to a $Z_2$-invariant neutrino mass matrix
defined at a high scale.
We discuss a specific model for this scenario.
In the concluding section~5
we make a comparison of the predictions
for $\ue3$ and $\cos{2 \theta_{23}}$
obtained within various frameworks.

\section{Vanishing $U_{e3}$ from a class of $Z_2$ symmetries}

The neutrino masses and lepton mixing are completely determined
by the neutrino mass matrix
in the flavour basis---the basis
where the charged-lepton mass matrix
is diagonal---which we denote as $\mnu$.
In this section we look for effective symmetries of $\mnu$
which may lead to a vanishing $U_{e3}$.

One knows~\cite{emt} that
the lepton-number symmetry $L_e - L_\mu - L_\tau$ implies
($i$) a vanishing solar mass-squared difference $\solm$,
($ii$) a maximal solar mixing angle $\theta_{23}$, 
and ($iii$) a vanishing $U_{e3}$,
while it keeps the atmospheric mixing angle unconstrained;
one must introduce~\cite{sf}
a significant breaking of $L_e - L_\mu - L_\tau$
in order to correct the predictions ($i$) and ($ii$).
A better symmetry seems to be
the $\mu$--$\tau$ interchange symmetry~\cite{mutau},
which implies vanishing $U_{e3}$ and maximal $\theta_{23}$,
but leaves both the neutrino masses
and the solar mixing angle unconstrained;
this is consistent with the present experimental results.
The $\mu$--$\tau$ interchange symmetry can be physically realized
in a model based on the discrete non-Abelian group $D_4$~\cite{d41};
a variation of this model~\cite{d42}
keeps the prediction $U_{e3} = 0$
but leaves the atmospheric mixing angle arbitrary.
Recently, Low~\cite{low} has considered models wherein $\mnu$ has,
due to a discrete Abelian symmetry, a structure leading to $U_{e3} = 0$.
 
We now show that there exists
a class $Z_2 \left( \gamma, \alpha \right)$
of discrete symmetries of the $Z_2$ type
which encompasses all the models discussed above
and enforces a form of $\mnu$ leading to $U_{e3} = 0$.
This class is parametrized by an angle $\gamma$
($0 < \gamma < 2 \pi$)
and a phase $\alpha$
($0 \le \alpha < 2 \pi$).
The symmetry $Z_2 \left( \gamma, \alpha \right)$
is defined by the $3 \times 3$ matrix
\be
S \left( \gamma, \alpha \right) = \left( \matrix{
1 &  0 & 0 \cr
0 &  \cos{\gamma} & e^{- i \alpha} \sin{\gamma} \cr 
0 & e^{i \alpha} \sin{\gamma} & - \cos{\gamma} \cr}
\right).
\ee
This matrix is unitary;
indeed,
it satisfies
\begin{eqnarray}
\left[ S \left( \gamma, \alpha \right) \right]^2 &=& 1_{3 \times 3},
\label{uirte} \\
\left[ S \left( \gamma, \alpha \right) \right]^T &=&
\left[ S \left( \gamma, \alpha \right) \right]^\ast.
\label{uhygt}
\end{eqnarray}
Equation~(\ref{uirte}) means that $S \left( \gamma, \alpha \right)$
is a realization of the group $Z_2$.
We define the $Z_2 \left( \gamma, \alpha \right)$ invariance of $\mnu$
by
\be
\left[ S \left( \gamma, \alpha \right) \right]^T \mnu
S \left( \gamma, \alpha \right) = \mnu.
\label{sym}
\ee
If one writes
\be
\mnu = \left( \matrix{
\tilde X & \tilde A & \tilde B \cr 
\tilde A & \tilde C & \tilde D\cr 
\tilde B & \tilde D & \tilde E\cr}
\right),
\ee
where all the matrix elements are complex in general,
then equation~(\ref{sym}) is equivalent to
\be
\begin{array}{rcl}
{\displaystyle \frac{\tilde B}{\tilde A}
- e^{- i \alpha} \tan{\frac{\gamma}{2}}}
&=& 0,
\\*[2mm]
\left( e^{i \alpha} \tilde E - e^{- i \alpha} \tilde C \right)
\sin{\gamma} + 2 \tilde D \cos{\gamma} &=& 0.
\end{array}
\label{relation}
\ee

Let us first prove that
the $Z_2 \left( \gamma, \alpha \right)$ invariance of $\mnu$
implies $U_{e3} = 0$.
The matrix $S \left( \gamma, \alpha \right)$
has a unique eigenvalue $- 1$ corresponding to the eigenvector
\be
v = \left( \begin{array}{c}
0 \\*[1mm]
{\displaystyle \exp \left( - i \alpha / 2 \right)
\sin{\left( \gamma / 2 \right)}}
\\*[2mm]
{- \displaystyle \exp \left( i \alpha / 2 \right)
\cos{\left( \gamma / 2 \right)}}
\end{array} \right).
\label{ev}
\ee
Equation~(\ref{sym}),
together with $S \left( \gamma, \alpha \right) v = - v$,
imply that $\left[ S \left( \gamma, \alpha \right) \right]^T
\left( \mnu\, v \right) = - \left( \mnu\, v \right)$.
Then,
equation~(\ref{uhygt}),
together with the fact that
the eigenvalue $- 1$ of $S \left( \gamma, \alpha \right)$ is unique,
implies that $\mnu\, v \propto v^\ast$.
Now,
$\mnu$ determines the lepton mixing matrix---MNS matrix---$U$ according to
\be
\mnu= U^\ast {\rm diag} \left( m_1, m_2, m_3 \right)  U^\dagger,
\label{mnuf}
\ee
where $m_1$,
$m_2$,
and $m_3$ are the (real and non-negative) neutrino masses.
Thus,
if we write $U = \left( u_1, u_2, u_3 \right)$,
then the column vectors $u_j$ satisfy $\mnu u_j = m_j u_j^\ast$
for $j = 1, 2, 3$.
The fact that $\mnu\, v \propto v^\ast$ therefore means that,
apart from a phase factor,
$v$ is one of the columns of the MNS matrix,
hence $U_{e3} = 0$, q.e.d.

Let us next prove the converse of the above,
i.e.\ that $U_{e3} = 0$ implies that there is some angle $\gamma$
and phase $\alpha$ such that
$\mnu$ is $Z_2 \left( \gamma, \alpha \right)$-invariant.
If $U_{e3} = 0$
then $U$ may be parametrized by two angles $\vartheta_{1,2}$
and five phases $\chi_{1,2,3,4,5}$ as
\be
U = \left( \begin{array}{ccc}
e^{i \chi_1} \cos{\vartheta_1} & e^{i \chi_2} \sin{\vartheta_1} & 0 \\
- e^{i \chi_3} \sin{\vartheta_1} \cos{\vartheta_2} &
e^{i \left( \chi_2 + \chi_3 - \chi_1 \right)}
\cos{\vartheta_1} \cos{\vartheta_2} &
e^{i \chi_4} \sin{\vartheta_2} \\
e^{i \chi_5} \sin{\vartheta_1} \sin{\vartheta_2} &
- e^{i \left( \chi_2 + \chi_5 - \chi_1 \right)}
\cos{\vartheta_1} \sin{\vartheta_2} &
e^{i \left( \chi_4 + \chi_5 - \chi_3 \right)} \cos{\vartheta_2}
\end{array} \right).
\label{probeU}
\ee
When one computes $\mnu$ through equation~(\ref{mnuf})
one then finds that it satisfies equations~(\ref{relation})
with $\gamma / 2 = \vartheta_2$ and
$\alpha = \chi_5 - \chi_3 + \pi$,
q.e.d.

One has thus proved the \emph{equivalence} of $U_{e3} = 0$
with the existence of some angle $\gamma$ and phase $\alpha$
such that $\mnu$ is $Z_2 \left( \gamma, \alpha \right)$-invariant.

It should be stressed that $Z_2 \left( \gamma, \alpha \right)$
will not usually be a symmetry of the full model,
nor is it necessarily the remaining symmetry
of some larger symmetry operating at a high scale.
Some examples may help making this clear:
\begin{itemize}
\item
The $\mu$--$\tau$ interchange symmetry [13],
which corresponds to $\cos{\gamma} = 0,\ e^{i \alpha} \sin{\gamma} = 1$,
cannot be a symmetry of the full theory,
since the masses of the $\mu$ and $\tau$ charged leptons
are certainly different;
thus,
that symmetry must be broken in the charged-lepton mass matrix,
but that breaking must occur in such a way
that it remains unseen---at least at tree level---in the form
of $\mnu$.
Moreover,
the $\mu$--$\tau$ interchange symmetry predicts
$\cos{2 \theta_{23}} = 0$ together with $U_{e3} = 0$.
\item
Many models based on $\bar L = L_e - L_\mu - L_\tau$ lead to [19]
\begin{equation}
\mnu = \left( \begin{array}{ccc}
x & y & r y \\ y & z & r z \\ r y & r z & r^2 z
\end{array} \right).
\label{lbar}
\end{equation}
In this case
$\cos{\gamma} = \left( 1 - |r|^2 \right) / \left( 1 + |r|^2 \right)$
and $e^{i \alpha} \sin{\gamma} = 2 r^\ast / \left( 1 + |r|^2 \right)$.
The symmetry $Z_2 \left( \gamma, \alpha \right)$
is not a subgroup of the original $\bar L$ symmetry,
rather it occurs accidentally
as a consequence of the specific particle content of the models
and of the particular way in which $\bar L$ is softly broken.
The mass matrix in equation~(\ref{lbar}) predicts $m_3 = 0$
together with $U_{e3} = 0$.
\item
The softly-broken $D_4$ model [16] has
\begin{equation}
\mnu^{-1} = \left( \begin{array}{ccc}
x & y & t \\ y & z & 0 \\ t & 0 & z
\end{array} \right),
\label{dfour}
\end{equation}
together with the condition $\arg{y^2} = \arg{t^2}$.
In this case
$\cos{\gamma} = \left( y^2 - t^2 \right) / \left( y^2 + t^2 \right)$
and $e^{i \alpha} \sin{\gamma} = 2 y t / \left( y^2 + t^2 \right)$.
The fact that the $\left( \mu, \tau \right)$ matrix element
of $\mnu^{-1}$ is zero,
and the fact that its $\left( \mu, \mu \right)$
and $\left( \tau, \tau \right)$ matrix elements remain equal,
are just reflections of the limited particle content
used to break the original $D_4$ symmetry softly.
\end{itemize}
Thus,
the symmetry $Z_2 \left( \gamma, \alpha \right)$ may be fundamental,
effective,
or accidental,
depending on the specific model at hand.

Considering equation~(\ref{probeU}) more carefully
one notices that the phase $\alpha = \chi_5 - \chi_3 + \pi$
is physically meaningless,
since it can be removed through a rephasing of the charged-lepton fields.
Let us then set $\alpha = 0$.
In that case,
the $\mnu$ satisfying equations~(\ref{relation})
can be written in the form
\be
\label{f1}
\mnu = \left( \matrix{
X &
\sqrt{2} A \cos{\left( \gamma / 2 \right)} &
\sqrt{2} A \sin{\left( \gamma / 2 \right)} \cr 
\sqrt{2} A \cos{\left( \gamma / 2 \right)} &
B + C \cos{\gamma} &
C \sin{\gamma} \cr 
\sqrt{2} A \sin{\left( \gamma / 2 \right)} &
C \sin{\gamma} &
B - C \cos{\gamma} \cr}
\right).
\ee
The eigenvalue corresponding to the eigenvector
in equation~(\ref{ev}) is $B - C$.

Specific choices of the parameters in equation~(\ref{f1})
give different models.
The model with $B = C = X = 0$
corresponds to $L_e - L_\mu - L_\tau$ symmetry~\cite{emt}.
The model with $\gamma = \pi / 2$
corresponds to $\mu$--$\tau$ interchange symmetry~\cite{mutau}.
The $D_4$ model in~\cite{d42}
has $\tilde X = \tilde A^2 / \tilde D$.
Likewise,
various models in~\cite{low} can be shown to have a $\mnu$
which is formally identical to the matrix in equation~(\ref{f1}).

In this paper we modify the standard parametrization for $U$
by multiplying its third row by $-1$, i.e. we use
\be
U = \left( \matrix{
c_{13} c_{12} &
c_{13} s_{12} &
s_{13} e^{-i \delta} \cr
- c_{23} s_{12} - s_{23} s_{13} c_{12} e^{i \delta} &
c_{23} c_{12} - s_{23} s_{13} s_{12} e^{i \delta} &
s_{23} c_{13} \cr
-s_{23} s_{12} + c_{23} s_{13} c_{12} e^{i \delta} &
 s_{23} c_{12} + c_{23} s_{13} s_{12} e^{i \delta} &
-c_{23} c_{13} \cr}
\right)
\times {\rm diag} \left( e^{i \rho},\ e^{i \sigma},\ 1 \right).
\label{parametrization}
\ee
Then, if we let $U_{e3} = s_{13} e^{-i \delta} = 0$,
equation~(\ref{mnuf}) reduces to equation~(\ref{f1})
with $\gamma / 2 = \theta_{23}$ and
\be
\label{abd2}
\begin{array}{rcl}
X &=& c_{12}^2 m_1 e^{- 2 i \rho}
+ s_{12}^2 m_2 e^{- 2 i \sigma},
\\*[2mm]
A &=& - {\displaystyle \frac{c_{12} s_{12}}{\sqrt{2}}}
\left( m_1 e^{- 2 i \rho}
- m_2 e^{- 2 i \sigma} \right),
\\*[3mm]
B &=& {\displaystyle \frac{1}{2}} \left(
s_{12}^2 m_1 e^{- 2 i \rho}
+ c_{12}^2 m_2 e^{- 2 i \sigma}
+m_3 \right),
\\*[3mm]
C &=& {\displaystyle \frac{1}{2}} \left(
s_{12}^2 m_1 e^{- 2 i \rho}
+ c_{12}^2 m_2 e^{- 2 i \sigma}
- m_3 \right).
\end{array}
\ee

\section{Non-zero $U_{e3}$,  $\cos 2\theta_{23}$
from $Z_{2}$ breaking}

Models with $U_{e3} = 0$ can be divided in two different categories:
\begin{itemize}
\item Those in which the solar scale also vanishes,
along with $U_{e3}$.
These are obtained by setting $m_1 = m_2$ in equations~(\ref{abd2}).
In these models,
the perturbation which generates the solar scale 
can be expected to also generate $U_{e3}$,
and one may find \cite{as,gou} correlations between them.
\item Models in which the solar scale
is present already at the zeroth order.
These are represented by equation~(\ref{f1})
without additional restrictions on its parameters,
except possibly $\gamma = \pi/4$.
\end{itemize}
We consider here the more general second category,
but fix $\gamma = \pi/4$,
i.e.\ we consider models with vanishing $U_{e3}$
and $\cos{2 \theta_{23}}$. 
$\mnu$ can be explicitly written in this case as
\be 
\mnu = U_0^\ast {\rm diag} \left( m_1, m_2, m_3 \right) U_0^\dagger,
\ee
where $U_0$ is obtained from equation~(\ref{parametrization})
by setting $s_{13} = 0$ and $\theta_{23} = \pi / 4$.
One then has
\be
\label{mnu0f}
\mnu = \left( \matrix{
X & A & A \cr
A & B & C \cr
A & C & B \cr}
\right).
\ee

Consider a general perturbation $\delta \mnu$
to equation~(\ref{mnu0f}).
The matrix $\delta \mnu$ is a general complex symmetric matrix,
but part of it can be absorbed through a redefinition
of the parameters in equation~(\ref{mnu0f}).
The remaining part can be written,
without loss of generality,
as
\be
\delta \mnu = \left( \matrix{
0 & \e_1 & - \e_1 \cr
\e_1 & \e_2 & 0 \cr
- \e_1 & 0 & - \e_2 \cr}
\right).
\label{deltm}
\ee
The perturbation is controlled by two parameters,
$\e_1$ and $\e_2$,
which are complex and model-dependent.
We want to study their effects perturbatively,
i.e.\ we want to assume $\e_1$ and $\e_2$ to be small.
This smallness can be quantified by saying
either that they are smaller than the largest element in $\mnu$,
or that the perturbation to a given matrix element
of $\mnu$ is smaller than the element itself.
We adopt the latter alternative and define
two dimensionless parameters:
\be
\e_1 \equiv \epsilon A, \quad \e_2 \equiv \epsilon^\prime B.
\ee
Thus, we have the neutrino mass matrix with $Z_2$ breaking as follows:
\be
\mnu = \left( \matrix{
X & A \left( 1 + \e \right) & A \left( 1 - \e \right) \cr
A \left( 1 + \e \right) & B \left( 1 + \e^\prime \right) & C \cr
A \left( 1 - \e \right) & C & B \left( 1 - \e^\prime \right) \cr}
\right) \ ,
\ee
\noindent where  we shall assume $\epsilon$ and $\epsilon^\prime$ to be small,
$\left| \e \right|, \left| \e^\prime \right| \ll 1$. 

One finds that,
to first order in $\e$ and $\e^\prime$,
the only effect of the $\delta \mnu$ in equation~(\ref{deltm})
is to generate non-zero $U_{e3}$ and $\cos{2 \theta_{23}}$.
The neutrino masses,
as well as the solar angle,
do not receive any corrections.
$U_{e3}$ and $\cos{2 \theta_{23}}$ are of the same order
as $\e$ and $\e^\prime$.
Define
\begin{eqnarray}
\hat m_1 &\equiv& m_1 e^{- 2 i \rho},
\\
\hat m_2 &\equiv& m_2 e^{- 2 i \sigma},
\end{eqnarray}
and
\begin{eqnarray}
\bar \e &\equiv& \left( \hat m_1 - \hat m_2 \right) \e,
\\
\bar \e^\prime &\equiv& \frac
{\hat m_1 s_{12}^2 + \hat m_2 c_{12}^2 + m_3}{2}\, \e^\prime.
\end{eqnarray}
Then, we get
\begin{eqnarray}
\label{spue3}
U_{e3} &=&
\frac{s_{12} c_{12}}{m_3^2 - m_2^2} \left(
\bar \e s_{12}^2 \hat m_2^\ast
+ \bar \e^\ast s_{12}^2 m_3
- \bar \e^\prime \hat m_2^\ast
- {\bar \e^\prime}^\ast m_3
\right)
\nonumber \\ & &
+ \frac{s_{12} c_{12}}{m_3^2 - m_1^2} \left(
\bar \e c_{12}^2 \hat m_1^\ast + \bar \e^\ast c_{12}^2 m_3
+ \bar \e^\prime \hat m_1^\ast
+ {\bar \e^\prime}^\ast m_3
\right),
\\
\label{spcos}
\cos{2 \theta_{23}} &=&
{\rm Re} \left\{
\frac{2 c_{12}^2}{m_3^2-m_2^2}
\left( \bar \e s_{12}^2 - \bar \e^\prime \right)
\left( \hat m_2 + m_3 \right)^\ast
- \frac{2 s_{12}^2}{m_3^2 - m_1^2}
\left( \bar \e c_{12}^2 + \bar \e^\prime \right)
\left( \hat m_1 + m_3 \right)^\ast
\right\}.
\end{eqnarray}

The meaningful phases in $\mnu$ are the ones of 
rephasing-invariant quartets.
Since $\mnu$ is symmetric,
there are three such phases which are linearly independent.
(Correspondingly,
there are three physical phases in the MNS matrix:
$\delta$,
$2 \rho$,
and $2 \sigma$.)
One easily sees that,
in the first-order approximation in $\e$ and $\e^\prime$,
the imaginary parts of those two small parameters
are meaningless when taken separately;
only $\mathrm{Im} \left( 2 \e - \e^\prime \right)$
is physically meaningful to this order.
Indeed,
one can manipulate equations~(\ref{spue3}) and~(\ref{spcos})
to obtain
\begin{eqnarray}
\cos{2 \theta_{23}} &=&
\left\{ \frac{c_{12}^2}{m_2^2 - m_3^2} \left[
m_3^2 + c_{12}^2 m_2^2 + s_{12}^2 {\rm Re}
\left( \hat m_1 \hat m_2^\ast + \hat m_1 m_3 \right)
+ \left( 1 + c_{12}^2 \right) {\rm Re}
\left( \hat m_2 m_3 \right) \right]
\right. \nonumber \\ & & \left. +
\frac{s_{12}^2}{m_1^2 - m_3^2} \left[
m_3^2 + s_{12}^2 m_1^2 + c_{12}^2 {\rm Re}
\left( \hat m_2 \hat m_1^\ast + \hat m_2 m_3 \right)
+ \left( 1 + s_{12}^2 \right){\rm Re}
\left( \hat m_1 m_3 \right) \right]
\right\} {\rm Re}\, \e^\prime
\nonumber \\ & &
+ 2 c_{12}^2 s_{12}^2
\left[
\frac{m_2^2 - {\rm Re} \left( \hat m_1 \hat m_2^\ast
+ \hat m_1 m_3 - \hat m_2 m_3 \right)}
{m_2^2 - m_3^2}
\right. \nonumber \\ & & \left.
+
\frac{m_1^2 - {\rm Re} \left( \hat m_1 \hat m_2^\ast
- \hat m_1 m_3 + \hat m_2 m_3 \right)}
{m_1^2 - m_3^2}
\right] {\rm Re}\, \e
\nonumber \\ & & + \frac{c_{12}^2 s_{12}^2 \left( m_1^2 - m_2^2 \right)
{\rm Im} \left( \hat m_1 \hat m_2^\ast + \hat m_1 m_3
+ \hat m_2^\ast m_3 \right)}
{\left( m_3^2 - m_1^2 \right) \left( m_3^2 - m_2^2 \right)}\,
{\rm Im} \left( 2 \e - \e^\prime \right),
\label{C1}
\end{eqnarray}
\begin{eqnarray}
\frac{U_{e3}}{c_{12} s_{12}} &=&
\frac{1}{2} \left\{ \frac{1}{m_2^2 - m_3^2}
\left[ s_{12}^2 \hat m_1 \hat m_2^\ast
+ s_{12}^2 \hat m_1^\ast m_3
+ \left( 1 + c_{12}^2 \right) \hat m_2^\ast m_3
+ m_3^2 + c_{12}^2 m_2^2 \right]
\right. \nonumber \\ & & \left.
+ \frac{1}{m_3^2 - m_1^2}
\left[ c_{12}^2 \hat m_1^\ast \hat m_2
+ c_{12}^2 \hat m_2^\ast m_3
+ \left( 1 + s_{12}^2 \right) \hat m_1^\ast m_3
+ m_3^2 + s_{12}^2 m_1^2 \right]
\right\} {\rm Re}\, \e^\prime
\nonumber \\ & &
+ \left[
\frac{s_{12}^2}{m_2^2 - m_3^2}
\left( m_2^2 - \hat m_1 \hat m_2^\ast - \hat m_1^\ast m_3
+ \hat m_2^\ast m_3 \right)
\right. \nonumber \\ & & \left.
+ \frac{c_{12}^2}{m_3^2 - m_1^2}
\left( m_1^2 - \hat m_1^\ast \hat m_2 + \hat m_1^\ast m_3
- \hat m_2^\ast m_3 \right)
\right]
{\rm Re}\, \e
\nonumber \\ & &
+ \frac{i}{2} \left[
\frac{s_{12}^2}{m_2^2 - m_3^2}
\left( m_2^2 - \hat m_1 \hat m_2^\ast + \hat m_1^\ast m_3
- \hat m_2^\ast m_3 \right)
\right. \nonumber \\ & & \left.
+ \frac{c_{12}^2}{m_3^2 - m_1^2}
\left( m_1^2 - \hat m_1^\ast \hat m_2 - \hat m_1^\ast m_3
+ \hat m_2^\ast m_3 \right)
\right]
{\rm Im} \left( 2 \e - \e^\prime \right).
\label{S1}
\end{eqnarray}

The induced values of $\ue3$ and $|\cos 2\theta_{23}|$ are strongly
correlated to neutrino mass hierarchies. This makes it possible to draw
some general conclusions even if we do not know the magnitudes of
$\e,\e'$.
In Table 1  we give expressions and values for
$\ue3$ and $|\cos 2\theta_{23}|$ in case of the 
hierarchical ($m_1<m_2<m_3$), inverted ($m_1\approx m_2\sim \sqrt{\atm}\gg
m_3$) and  quasi-degenerate neutrino spectrum. CP conservation is
assumed but we distinguish two different cases (a) the Dirac solar pair
corresponding to $\sigma=\rho=0$ and the Pseudo-Dirac solar pair
with\footnote{The physically different case with $\rho=0,\sigma=\pi/2$ 
has similar results.} $\rho=\pi/2,\sigma=0$. We have also given approximate
values in some cases assuming the common degenerate mass $m\sim 0.3 ~\eV$.

It follows from the Table 1 and equations~(\ref{C1},\ref{S1}) that:
\begin{itemize}

\item The first order contribution to $U_{e3}$ given in
equation~(\ref{S1}) 
vanish identically if $\hat{m}_1=\hat{m}_2$. As a consequence of this,
$U_{e3}$ gets
suppressed by a factor $\ord(\frac{\solm}{\atm})$ for the
inverted or quasi-degenerate spectrum with $\rho=\sigma=0$. Similar
suppression also occurs in case of the normal neutrino mass hierarchy
even when $\rho\not =\sigma$. $U_{e3}$ need not be suppressed in other
cases and can be large.

\item In contrast to $U_{e3}$, $\cos 2\theta_{23}$ is almost as large as 
$\e,\e'$ if neutrino mass spectrum is normal or inverted. It gets enhanced
compared to these parameters if the spectrum is quasi-degenerate.
\item
In case of the quasi-degenerate spectrum,
both $|\cos2\theta_{23}|$ and  $\ue3$  can become quite large and reach
the present experimental limits.
Especially, the enhancement factors  are large
in case of the pseudo-Dirac solar pair ($\rho=\pi/2$,\ $\sigma=0$).
$U_{e3}$ and $\cos 2 \theta_{23}$ are in fact proportional to each other
in this particular case. The parameters $\e,\e'$ are constrained to be
lower than $10^{-2}$ for the quasi-degenerate spectrum.
\end{itemize}
\begin{table}{}
\begin{tabular}{|l|l|}
\hline
{\bf Normal Hierarchy} &
$|U_{e3}|\approx c_{12} s_{12}\sqrt{\soa} (\e+\frac{\e'}{2})\approx 
0.09 (\e+\frac{\e'}{2})$\\[.3cm]
$m_1\ll  m_2;m_2^2\approx \solm;m_3^2\approx \atm$ & $|\cos 2 \theta_{23}|\approx
\e'$\\[.3cm]
\hline
{\bf Inverted Hierarchy}&$|U_{e3}|\approx {\solm\over
2\atm}s_{12}c_{12}(\e-\frac{\e'}{2})\approx 0.009(\e-\frac{\e'}{2})$
\\[.3cm]
$\sigma=0;\rho=0$&$|\cos 2 \theta_{23}|\approx \e'$\\[.25cm] 
\hline
$\sigma=0;\rho=\pi/2$& $|U_{e3}|\approx \frac{1}{2} \sin 4
\theta_{12} (\e-\frac{\e'}{2})\approx 0.4(\e-\frac{\e'}{2}) $\\[.3cm]
&$|\cos 2 \theta_{23}|\approx 2 (\e \sin^2 2\theta_{12}+\frac{\e'}{2} 
\cos^2
2\theta_{12})$\\ [.3cm]
\hline
{\bf Quasi-Degenerate}&$|U_{e3}|\approx  2 \e'c_{12}s_{12} {m^2\over \atm} 
\soa\approx 1.6 \e$\\ [.3cm]
$\sigma=0;\rho=0$
& $|\cos 2 \theta_{23}|\approx 4 \frac{m^2}{\atm} \e'\approx 180 \e'$\\[.3cm] 
\hline
$\sigma=0;\rho=\pi/2$& $|U_{e3}|\approx 4{m^2\over \atm}c_{12} s_{12}(\e
s_{12}^2+\frac{\e'}{2} c_{12}^2)\approx 81(\e
s_{12}^2+\frac{\e'}{2} c_{12}^2)$\\[.3cm] 
& $|\cos 2 \theta_{23}|\approx 8 {m^2\over \atm} c_{12}^2 
(\e s_{12}^2+\frac{\e'}{2} c_{12}^2)\approx 259 (\e
s_{12}^2+\frac{\e'}{2} c_{12}^2)$\\[.3cm] 
\hline
\end{tabular}  
\caption{Leading order predictions for $\ue3,\ |\cos 2\theta_{23}|$ in case of
different neutrino mass hierarchies with CP conservation. 
The numerical estimates are based on the best fit values of 
neutrino parameters and the quasi-degenerate mass $m=0.3\eV$.}
\end{table}
\vskip 0.0cm
The perturbative expressions given above may not be reliable 
for some values of $\e,\e'$ due to large
enhancement factor of $\ord({m^2\over \atm})$ and one should do 
a numerical analysis. We now discuss results of such analysis
in various circumstances. Scattered plots of the predicted values
for  $|\cos 2\theta_{23}|$ and  $\ue3$ are given in Figure 1 in the case
of normal neutrino mass hierarchy. 
CP conservation ($\rho=\sigma=0$, real $\e,\e'$) is assumed. 
Neutrino masses and  $\theta_{12}$ do not
receive any corrections at $O(\e,\e')$ and hence do not appreciablly
change by perturbations. We therefore randomly varied these input
parameters in the experimentally allowed regions. $m_1$ was varied up to
$m_2$. On the other hand,  $\e,\e'$ are unknown 
unless the symmetry breaking is specified, so these are 
varied randomly in the range $-0.3\sim 0.3$ with the condition that the output
parameters should lie in the $90\%$ CL limit \cite{Kamland,lisi}:
\begin{eqnarray}
 0.33 \le  \tan^2 \theta_{\rm sun} \le 0.49  \ , \ \
&&7.7\times 10^{-5} \le  \Delta_{\rm sun} \le
8.8\times 10^{-5}~\rm{eV}^2,
 \quad 90\% C.L. \ ,\nonumber \\
0.92 \le \sin^2 2\theta_{\rm atm}  \ , \ \
&& 1.5\times 10^{-3} \le  \Delta_{\rm atm} \le 3.4 \times 10^{-3}~
 \rm{eV}^2 ,  \quad 90\% C.L. \ .
\label{data}
\end{eqnarray} 

\begin{figure}
\begin{center}
\epsfxsize=12.0 cm
\epsfbox{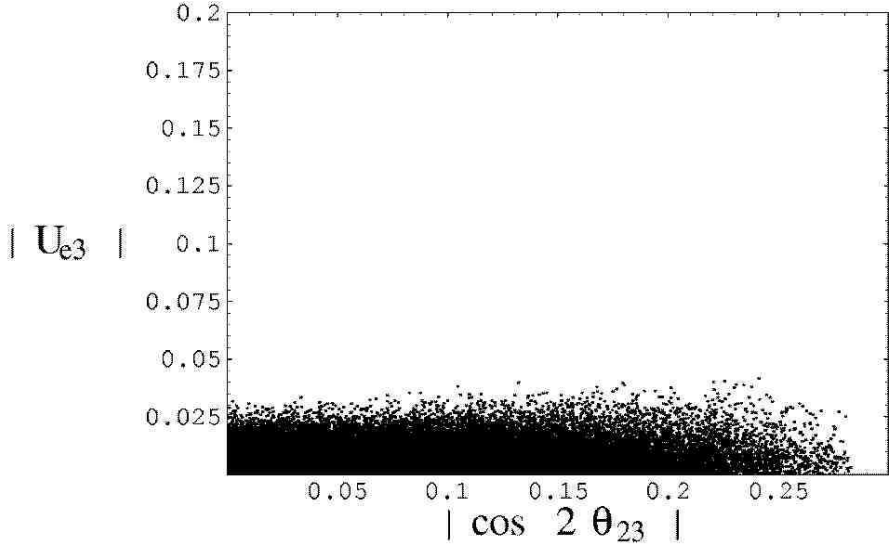}
\end{center}
\vspace*{0.cm}
\caption{The scattered plots showing the allowed values of $|\cos 2
\theta_{23}|$ and $\ue3$ in case of the normal neutrino mass hierarchy.
$\e,\e'$ are randomly varied in the range $-0.3\sim 0.3$.
The Majorana phases are chosen as
$\rho=0, \ \sigma=0$.}
\end{figure}%
The $\ue3$ is forced to be small less than $0.025$, in Figure 1 as would
be expected from the foregoing discussion. The value $\sim 0.025$ at the
upper end arises
from the (assumed) bound $|\e|,|\e'|\leq 0.3$. Since $\ue3$ is
proportional
to $\e,\e'$, it increases if  the bound on $\e,\e'$ is loosened.
However,  $|\e|\leq  0.3$ is a reasonable bound due to assume if
$Z_2$ breaking is perturbative. On the other hand,   
$|\cos 2\theta_{23}|$  can assume large values as seen from Figure 1.
The present bound $\sin^2 2\theta_{23}>0.92$ from the atmospheric
experiments gets translated to $|\cos 2\theta_{23}|<0.28$ which constrains
 $|\e'| \leq 0.2$ in our analyses.

The non-maximal value for $\theta_{23}$ gives rise to interesting physical
effects such as excess of the e-like events in the atmospheric neutrino
data in the sub-GeV region \cite{sp}, different matter dependent
survival probabilities for the $\nu_{\mu}$ and the $\bar{\nu}_\mu$
\cite{probir}. These can be searched for in the future atmospheric \cite{Gon}
and the long baseline experiments. The values
$|\cos 2\theta_{23}|>0.1$ are expected to be probed in these experiments
\cite{lbase2}.  These values occur quite naturally
for a reasonably large range of parameters.
\begin{figure}
\begin{center}
\epsfxsize=12.0 cm
\epsfbox{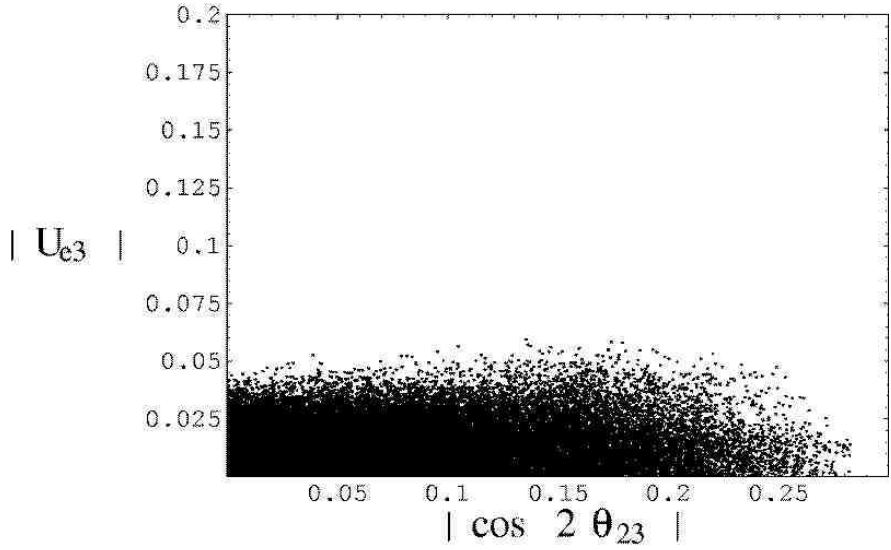}
\end{center}
\vspace*{0.0cm}
\caption{The allowed values  $|\cos 2 \theta_{23}|$ and $\ue3$ for 
$\rho=\pi/4, \ \sigma=0$ and the normal neutrino mass hierarchy. 
The other parameters are the same as in Fig. 1.}
\end{figure}%
\begin{figure}
\begin{center}
\epsfxsize=12.0 cm
\epsfbox{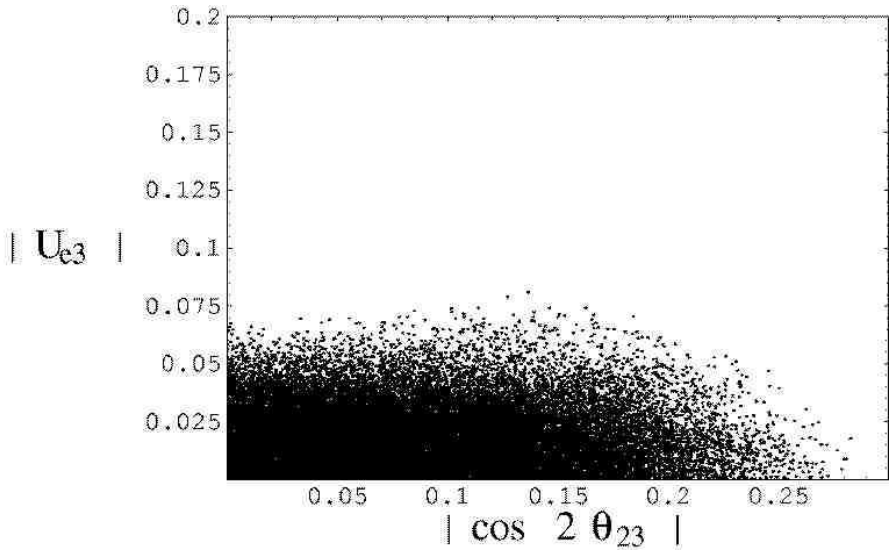}
\end{center}
\vspace*{0.0cm}
\caption{The allowed values  $|\cos 2 \theta_{23}|$ and $\ue3$ for 
$\rho=\pi/2, \ \sigma=0$ and the normal neutrino mass hierarchy. 
The other parameters are the same as in Fig. 1.}
\end{figure}%
In order to find the phase dependence of our results, we show the results
 in the cases of ($\rho=\pi/4, \ \sigma=0$) and ($\rho=\pi/2, \ \sigma=0$).
The phase dependence is found in the prediction of  $\ue3$, which 
 increases up to $0.075$. 

The region $\ue3>0.07$ is expected to be
probed in the long baseline experiments with the conventional or 
super beams \cite{lbase} and in the reactor experiments \cite{dchooz}. 
The smaller values for $\ue3 \sim 0.025$ can be
reached only at the neutrino factory \cite{nf}. Most of the region
displayed in Figs. 1-3  therefore seem inaccessible to 
the near future neutrino experiments aimed at searching for $\ue3$.

\begin{figure}
\begin{center}
\epsfxsize=12.0 cm
\epsfbox{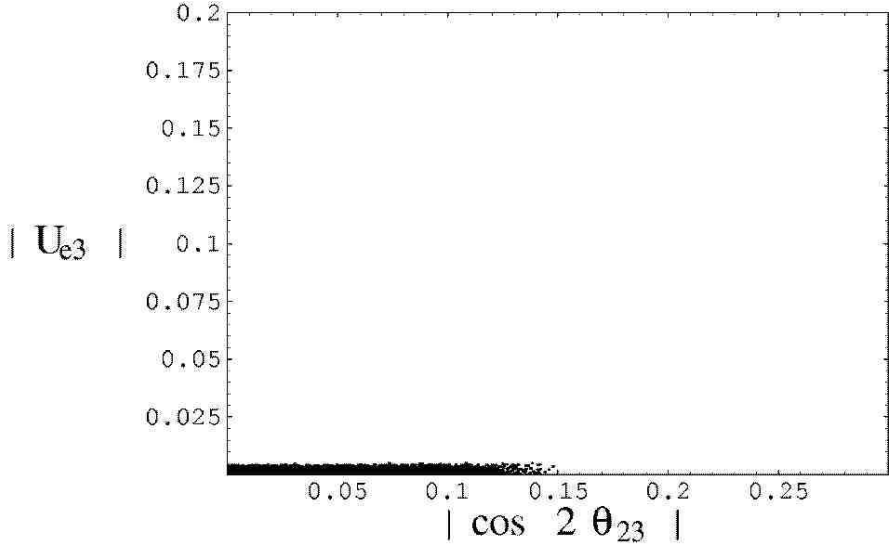}
\end{center}
\vspace*{.0cm}
\caption{The allowed values of $|\cos 2 \theta_{23}|$ and $\ue3$ for 
$\rho=0, \ \sigma=0$ in case of  the inverted neutrino mass hierarchy. 
The $\e,\e'$ are varied randomly  in the range $-0.3\sim 0.3$
while $m_3$ is varied up to $10^{-2}\eV$.}
\end{figure}%
Scattered plots for 
the predicted values for $\ue3$ and $|\cos 2\theta_{23}|$ are given in Fig. 4 
in  case of the inverted  hierarchy of the neutrino masses.
The value of  $\ue3$ is even more suppressed compared to the corresponding
case displayed in Fig. 1.
This suppression is due to the strong cancellation
between $m_1$ and $m_2$, which is seen in Table 1.
However, the Majorana phases spoil  this cancellation, and so $\ue3$ 
could be larger as seen in Figs. 5 and 6, where the
two cases ($\rho=\pi/4, \ \sigma=0$) and ($\rho=\pi/2, \ \sigma=0$) are
displayed respectively.
Thus, the effect of the Majorana phases is very important 
in the inverted hierarchy.
The isolated points in Fig. 6 follows from  the tuning of the parameters
$\e$ and $\e'$. Apart from this tuning, the allowed values of $\ue3$ are
moderate $\sim 0.1$ but will be explored 
in the future long baseline and reactor experiments.
\begin{figure}
\begin{center}
\epsfxsize=12.0 cm
\epsfbox{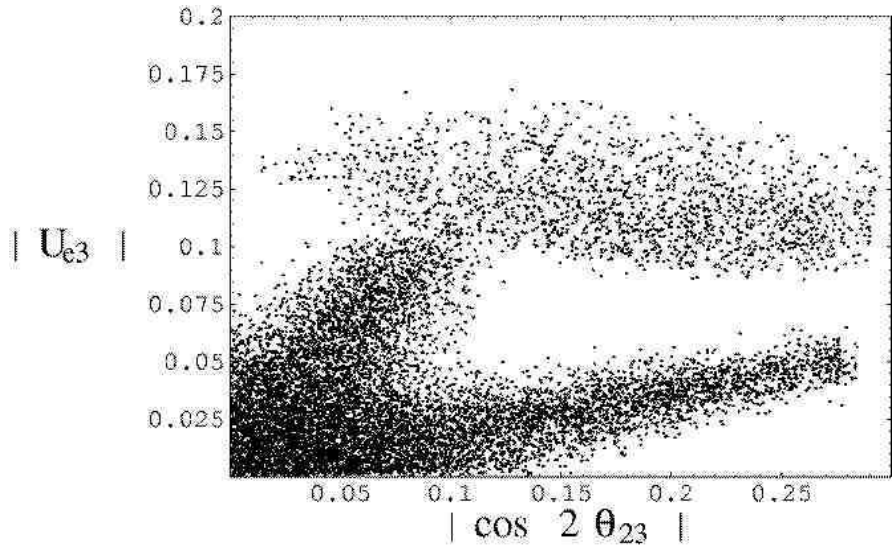}
\end{center}
\vspace*{.0cm}
\caption{The allowed values of $|\cos 2 \theta_{23}|$ and $\ue3$ 
for $\rho=\pi/4, \ \sigma=0$ 
in case of the inverted neutrino mass hierarchy.}
\end{figure}%
\begin{figure}
\begin{center}
\epsfxsize=12.0 cm
\epsfbox{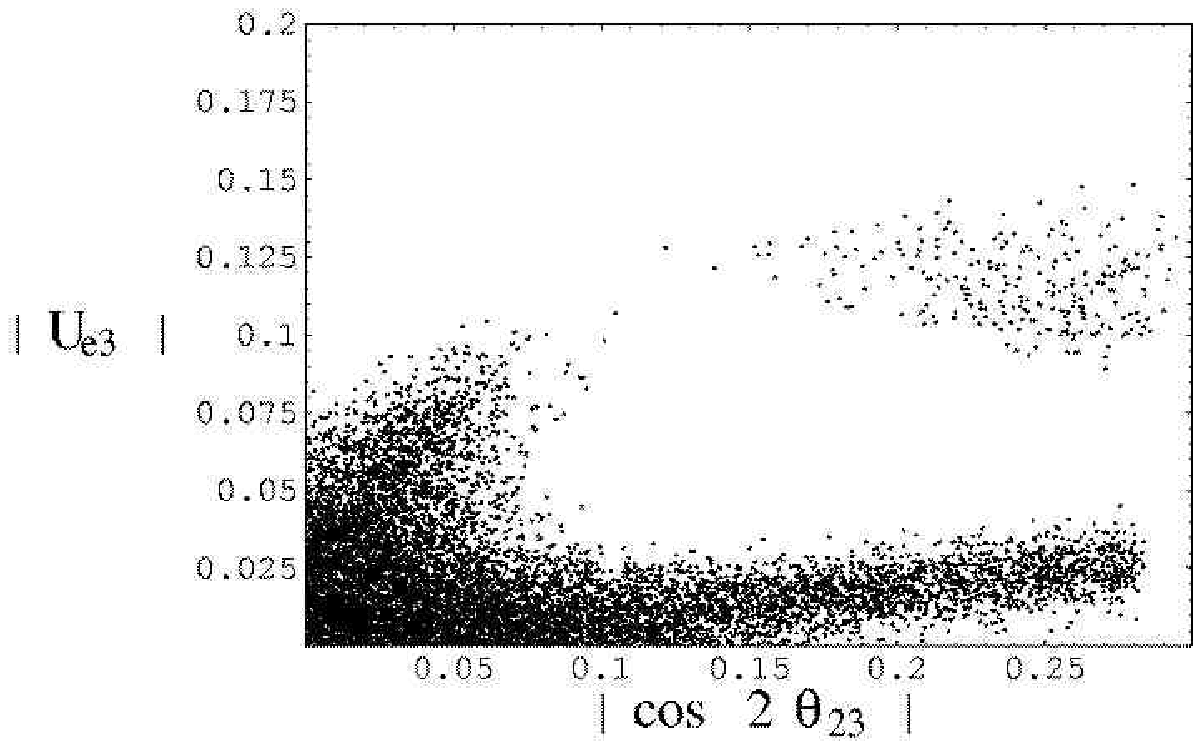}
\end{center}
\vspace*{.0cm}
\caption{The allowed values of $|\cos 2 \theta_{23}|$ and $\ue3$ 
for $\rho=\pi/2, \ \sigma=0$ 
in case of the inverted neutrino mass hierarchy.}
\end{figure}%
\begin{figure}
\begin{center}
\epsfxsize=12.0 cm
\epsfbox{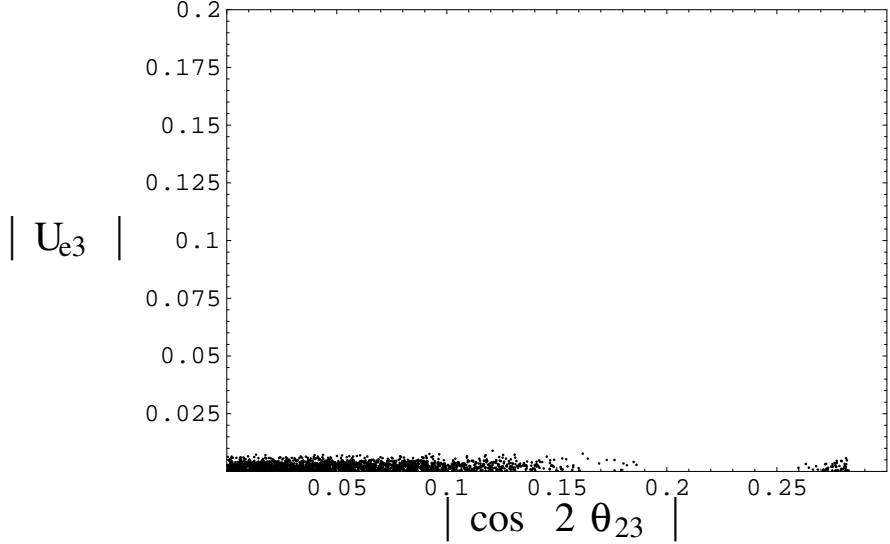}
\end{center}
\vspace*{.0cm}
\caption{The scattered plots of the allowed values of $|\cos 2
\theta_{23}|$ and 
$\ue3$  
with $|\e|\leq 0.3$ and $|\e'|\leq 0.03$
and the quasi-degenerate neutrino masses. The Majorana phases are chosen
as $\rho=0, \ \sigma=0$. The degenerate mass scale is fixed at
 $m=0.3\ \eV$.}
\end{figure}%

The parameter $\e'$ is
constrained strongly $|\e'|\leq 0.03$ in case of the quasi-degenerate
neutrino masses due to an enhancement factor $\ord(\frac{m^2}{\atm})$
present in this case, as seen in Table 1.
$|\e|$ is however not constrained as strongly and we take 
 $|\e|\leq 0.3$. The scattered plots for  the predicted values 
for $\ue3$ and $|\cos 2\theta_{23}|$ are given in Fig. 7. 
 The value of  $\ue3$ is  expected to be $O(0.01)$.
There are  partial  cancellations among contributions from $m_1$, $m_2$,
$m_3$ when $\rho=\sigma$.
However, different choice for the  Majorana phases spoil  this
cancellation and  $\ue3$ 
could be large as seen in Figs. 8 and 9,
which correspond to ($\rho=\pi/4,\ \sigma=0$) and ($\rho=\pi/2,\
\sigma=0$) respectively. It is found that $\ue3$ could increase to $0.1$
in these cases.
\begin{figure}
\begin{center}
\epsfxsize=12.0 cm
\epsfbox{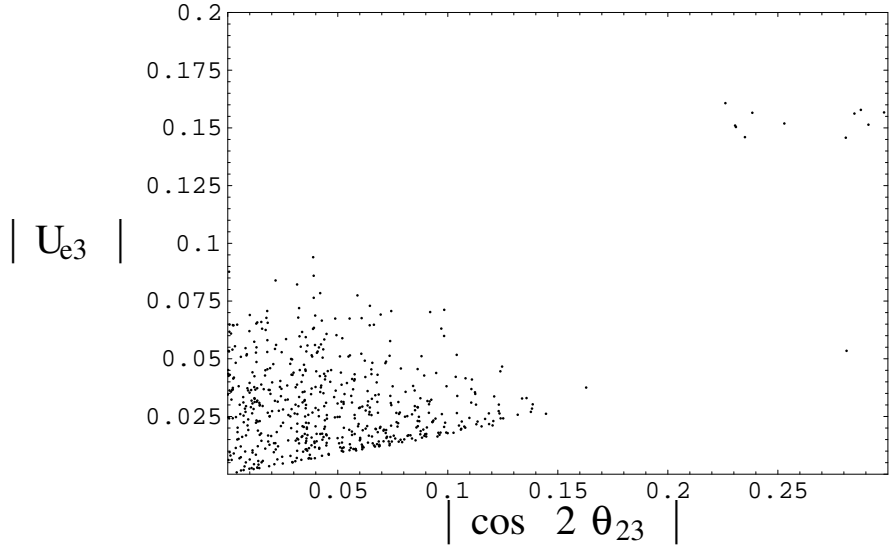}
\end{center}
\vspace*{.0cm}
\caption{The allowed values of $|\cos 2\theta_{23}|$ and 
$\ue3$  in the quasi-degenerate neutrino masses. 
The Majorana phases are chosen
as $\rho=\pi/4, \ \sigma=0$. The degenerate mass scale is fixed at
 $m=0.3\ \eV$.}
\end{figure}%
\begin{figure}
\begin{center}
\epsfxsize=12.0 cm
\epsfbox{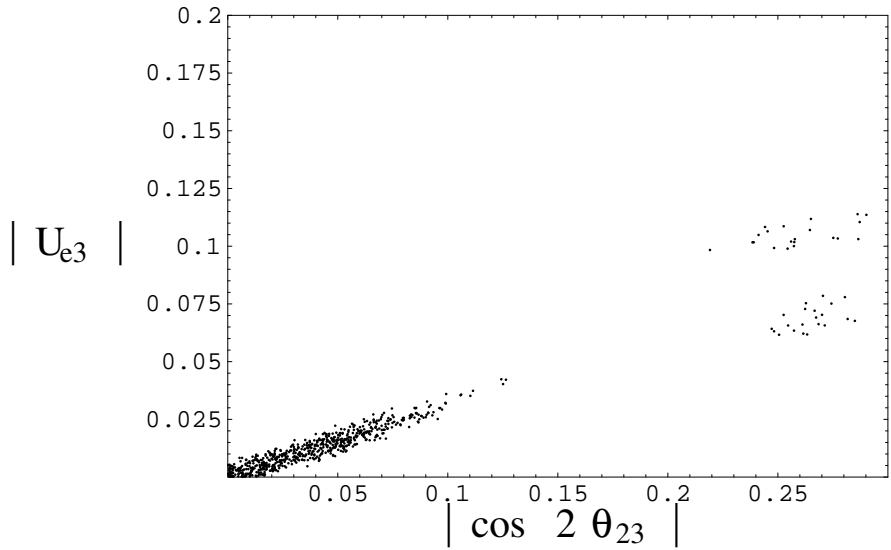}
\end{center}
\vspace*{.0cm}
\caption{The allowed values of $|\cos 2\theta_{23}|$ and 
$\ue3$  in the quasi-degenerate neutrino masses.
 The Majorana phases are chosen
as $\rho=\pi/2, \ \sigma=0$. The degenerate mass scale is fixed at
 $m=0.3\ \eV$.}
\end{figure}%

 In the above analyses, we fixed $\sigma=0$ because
 only the relative phase $\rho-\sigma$ is essential in determining the
 masses and mixing angles in the case of 
the hierarchical and inverted hierarchical neutrino masses.
However, $\sigma$ dependence is non-trivial for the degenerate masses.
We show the results for
($\rho=0, \ \sigma=\pi/2$)  and ($\rho=\pi/4, \ \sigma=\pi/2$) in
Fig. 10 and Fig. 11 respectively.
It is noted  that $\ue3$ could be as large as $0.2$ for the case 
$\rho=\pi/4, \ \sigma=\pi/2$ but values $\leq 0.1$ are more probable as
seen from the density of points. 
\begin{figure}
\begin{center}
\epsfxsize=12.0 cm
\epsfbox{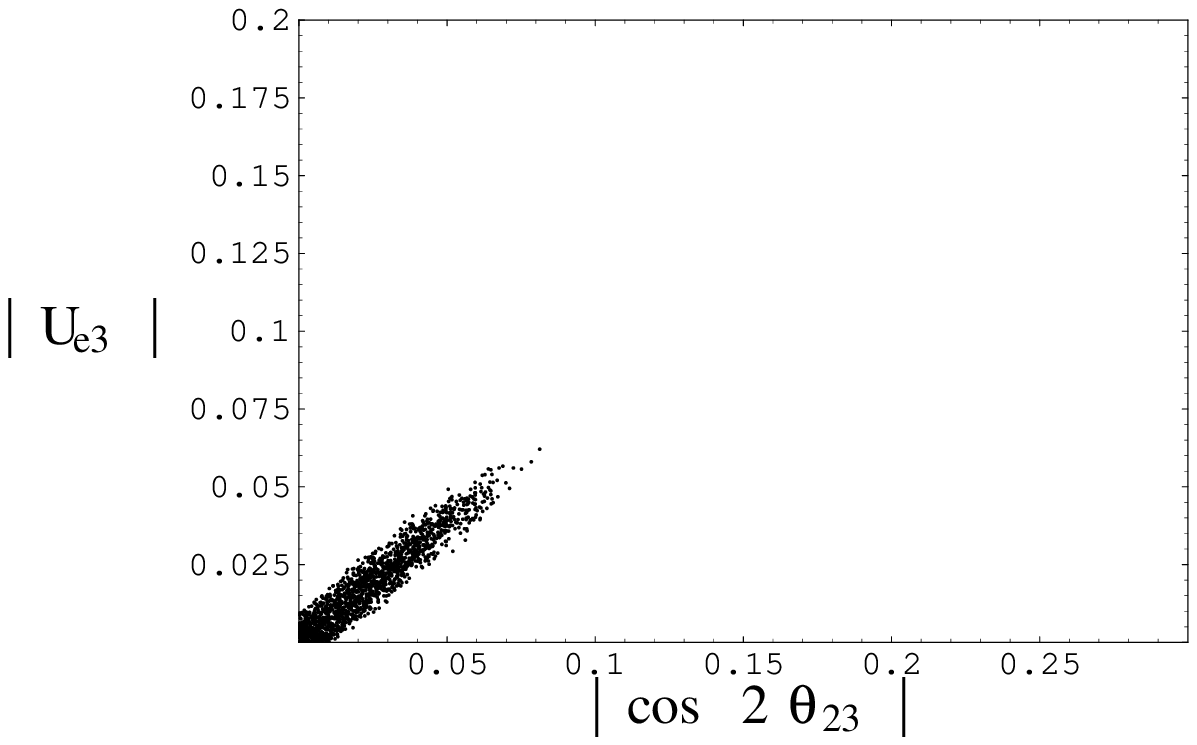}
\end{center}
\vspace*{1.0cm}
\caption{The allowed values of $|\cos 2\theta_{23}|$ and 
$\ue3$  in the quasi-degenerate neutrino masses. 
The Majorana phases are chosen
as $\rho=0, \ \sigma=\pi/2$. The degenerate mass scale is fixed at
 $m=0.3\ \eV$.}
\end{figure}%
\begin{figure}
\begin{center}
\epsfxsize=12.0 cm
\epsfbox{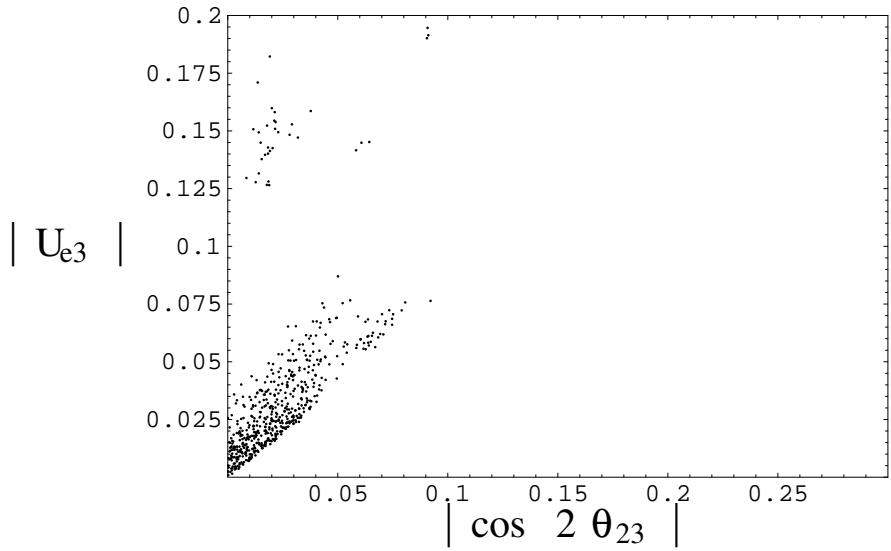}
\end{center}
\vspace*{1.0cm}
\caption{The allowed values of $|\cos 2\theta_{23}|$ and 
$\ue3$  in the quasi-degenerate neutrino masses.
 The Majorana phases are chosen
as $\rho=\pi/4, \ \sigma=\pi/2$. The degenerate mass scale is fixed at
 $m=0.3\ \eV$.}
\end{figure}%

Before ending this section, we wish to point out an interesting aspect of
this analysis. Since $U_{e3}$ is zero in the absence of the perturbation,
the CP violating Dirac phase $\delta$ relevant for neutrino oscillations
is
undefined at this stage. CP violation is present through the Majorana
phases $\rho$ and $\sigma$. Turning on perturbation leads to non-zero
$U_{e3}$ and also to a non-zero Dirac phase even if perturbation is real.
Moreover, $\delta$ generated this way can be large and independent of the
strength of perturbation parameters. This phenomenon  was noticed \cite{asn} 
in the specific case of the radiative generation of $U_{e3}$. 
This occurs here also for a more general perturbation. 

As an example, let us consider the limit $\e'=0$ and a real $\e$. Since
$U_{\mu 3}$ is almost maximal and real, $\delta$ is approximately given by
\be
\tan\delta\approx {m_1 m_2 \sin 2(\rho-\sigma)-m_3m_1 \sin 2\rho+m_2 m_3
\sin 2\sigma +\ord(\soa) Im({\mathcal Z}) \over  m_1^2 c_{12}^2-
 m_2^2  s_{12}^2-m_1 m_2 \cos 2\theta_{12}\cos 2(\rho-\sigma)+
 m_3 m_1 \cos 2\rho-
m_2 m_3 \cos 2 \sigma+\ord(\soa) Re({\mathcal Z})} ~,\ee
where ${\mathcal Z}\equiv (\hat{m}_2^* (\hat
{m}_1-\hat{m_2})+m_3(\hat{m}_1-\hat{m_2})^*)s_{12}^2$. 
It follows from above that irrespective of the specific mass hierarchy, 
the induced $\delta$ would be large if $\rho$ and $\sigma$ are large
and not finetuned.

\section{Radiatively generated $U_{e3}$ and $\cos 2\theta_{23}$}

The $\e,\e'$ were treated as independent parameters so far. They can be
related in specific models. We now consider one example which is based
on the electroweak breaking of the $Z_2$ symmetry in the MSSM.
We assume that neutrino masses are generated at some
high scale $M_X$ and the effective neutrino mass operator describing them
is  $Z_2$ symmetric with the result that $U_{e3}= \cos 2\theta_{23}=0$ 
at $M_X$. This symmetry is assumed to be broken spontaneously
in the Yukawa couplings of the charged leptons. This breaking would
radiatively induce non-zero $U_{e3}$ and $\cos 2\theta_{23}$ \cite{Ratz}.
 This can be
calculated by using the renormalization group equations (RGEs) of the
effective neutrino mass operator \cite{rg1,rg2,rg3}. 
These equations depend upon the 
detailed structure of the model below $M_X$. We assume here that theory
below $M_X$ is the MSSM and use the RGEs
derived in this case. Subsequently we will give an example 
which realizes our assumptions.

Integration of the RGEs allows us \cite{rg1,rg2,rg3} to relate the
neutrino mass matrix $\mnu (M_X)$ to the corresponding matrix at the low scale
which we identify here with the $Z$ mass $M_Z$:
\be \label{corrected}\mnu(M_Z) \approx I_g
I_t ~(I~ \mnu (M_X) ~I~)\ , \ee
where $I_{g,t}$ are calculable numbers depending on the gauge and
top quark Yukawa couplings. $I$ is a flavour dependent matrix
given by
\be
 I\approx \mathrm{diag}(1+\delta_e,1+\delta_\mu,1+\delta_\tau) \ ,
\ee
with 
\be \label{delta1}
\delta_\alpha\approx c\left({m_\alpha\over 4 \pi v }\right)^2 \ln{M_X\over
M_Z}~, 
\ee
where $c=\frac{3}{2},-\frac{1}{\cos^2\b}$ in case of the Standard
Model (SM) and  the Minimal Supersymmetric Standard Model (MSSM)
respectively \cite{rg1}. $v$ refers to the vacuum expectation value for
the SM Higgs doublet.

We have implicitly neglected possible threshold effects. Inclusion of
these effects would not modify the analysis if threshold effects
are flavour blind as would be approximately true \cite{brahma} in case of
the minimal supergravity scenario with universal boundary conditions.

$\mnu(M_X)$ is given by equation~(\ref{mnu0f}). From this we can write 
$\mnu (M_Z)$ as follows when
the muon and the electron Yukawa couplings are neglected:
\begin{eqnarray}
\mnu (M_Z)
&=& \left(\matrix{ X&A'&A'\cr
                       A'&B'&C'\cr
                       A'&C'&B'\cr}\right)
    +\left( \matrix{ 0&A' \e&-A'\e \cr
                       A' \e&B' \e'&0 \cr
                       -A'\e&0&-B'\e' \cr} \right)+O(\delta_\tau^2) ~,
\end{eqnarray}          
where
\be \label{abdrad}
C'=C(1+\dt)~~,~~A'=A(1+\frac{\dt}{2})~~,~~
B'=B(1+\dt)~~,~~\e=\frac{\e'}{2}=-\frac{\dt}{2} \ ,
\ee
and $A,B,C$ are defined in equation~(\ref{abd2}).
Note that $m_1$, $m_2$ and $m_3$ defined previously are no longer mass
eigenvalues
because of the changes $A\rightarrow A'$,  $B\rightarrow B'$ and
  $C\rightarrow C'$.
Using the above equations, we get from equation~(\ref{spue3})
\begin{eqnarray}\label{ue3rad}
U_{e3} \simeq&& \hskip -0.5cm -{ \dt s_{12}c_{12}\over2(
m_3^2-m_1^2)}\left[m_1^2+2 m_3 \hat{m_1}^*+m_3^2\right] 
 + { \dt s_{12}c_{12}\over 2 m_3^2-m_2^2}
\left[m_2^2+2 \hat{m_2}^* m_3+m_3^2 \right]\ , \nonumber \\ 
\cos 2 \theta_{23} \simeq&& \hskip -0.5cm  {\dt s_{12}^2\over
m_3^2-m_1^2}\left[m_1^2+2 m_3 \hat{m_1}^*+m_3^2\right]
+ {\dt c_{12}^2\over m_3^2-m_2^2}
\left[m_2^2+2 \hat{m_2}^* m_3+m_3^2 \right] \ .
\end{eqnarray}

It is easily seen that the effect of the radiative corrections is enhanced 
in the case of the  quasi-degenerate neutrino masses with opposite phase 
$|\rho-\sigma|=\pi/2$ as  previous works presented \cite{rg2,rg3}.
In the MSSM, the parameter $\dt$ is negative and its absolute value 
 can become quite large for large $\tan\b$, {\it e.g.}
for $\tan\b\sim 50,\  |\dt|\sim 0.075$. 
However, large $\tan\b$ is not favoured because
the renormalization of parameters $A,B,C$ as in equation~(\ref{abdrad}) also shifts
the value of the solar angle and solar mass compared to their values in
the $\dt \rightarrow 0$ limit. One now gets
\begin{equation} \label{solar}
\solm \cos 2\sola\approx \Delta_{21}\cos 2\theta_{12}+2 \dt |m_1 e^{-2 i
\rho} s_{12}^2+m_2 e^{-2 i \sigma} c_{12}^2|^2 ~. \ee
Here, $\Delta_{21}\equiv m_2^2-m_1^2$ and $\theta_{12}$ correspond to
the values of the solar scale and angle at $M_X$. The radiative
corrections add a negative contribution to $\solm \cos 2\sola$ 
in case of the MSSM and can spoil the LMA solution (which need positive
$\solm \cos 2\sola$) if $\Delta_{21}$ is small or $|\dt|$ is large. 
This provides a constraint on possible values of $\dt$ and consequently on 
$|U_{e3}|,|\cos 2\theta_{23}|$ that can be generated in the model. 
For example, requiring that the first term dominates over the
second term in equation~(\ref{solar}) implies 
\begin{eqnarray}
|\dt|\leq \left({\Delta_{21}\over 2 m^2 \cos 2\theta_{12}}\right)\approx
10^{-3}  ~ , 
\end{eqnarray}
where we assumed CP conservation, the quasi-degenerate spectrum, 
$\sigma=\pi/2;\ \rho=0$ , $m\approx 0.3\ \eV$,
and $\Delta_{21}\sim 8\times 10^{-5}$ eV$^2$. The values for $\ue3$
and $|\cos 2\theta_{23}|$ implied by the above constraint are quite small.
Notice however that one can loosen the bound on $\dt$ by choosing
significantly larger value $\Delta_{21}$ than $8\times 10^{-5} \eV^2$.
The cancellations between two terms in equation~(\ref{solar}) can still
lead to
physical solar scale. 

\begin{figure}
\begin{center}
\epsfxsize=12.0 cm
\epsfbox{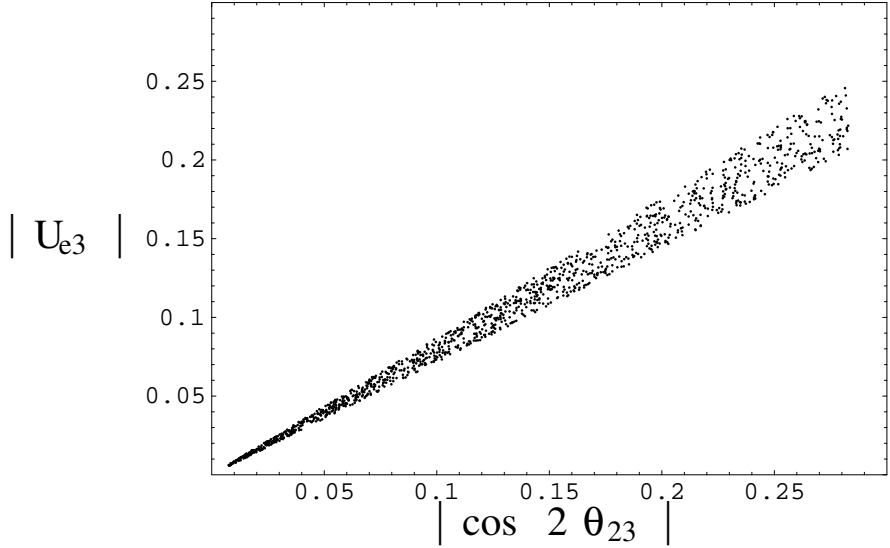}
\end{center}
\vspace*{0.0cm}
\caption{The scattered plots  of the allowed values of $|\cos 2
\theta_{23}|$ and $\ue3$  
in case of the radiatively broken $Z_2$ and the quasi-degenerate neutrino
masses $m=0.3{\rm eV}$. The Majorana phases are chosen as
$\rho=0, \ \sigma=\pi/2$.}
\end{figure}%
Results of the numerical analysis are shown in Fig. 12 in case of the
quasi-degenerate spectrum with $m=0.3 ~\eV;\sigma=\pi/2,\rho=0$. 
The $\theta_{12},\Delta_{21}$ and $\tan \beta$ at high scale are varied
randomly, then  the allowed choices which reproduce the parameters as in
equation~(\ref{data}) at the low scale are determined. 
Both $\ue3$ and $|\cos 2\theta_{23}|$ can reach their respective 
experimental bound. The near proportionality between the two can be
understood from their expressions given in Table 1. We find numerically
that $\tan\beta$ is constrained to be lower than $20$ in this case.
The forthcoming
experiments will be able to test this relationship between $\ue3$ and 
$|\cos 2\theta_{23}|$.
 It may be useful to note  our numerical results of  $\ue3$ 
in the cases of the normal-hierarchy  and inverted-one of the neutrino masses.
 In both cases, $\ue3$ reaches at most $0.025$.
These results are consistent with one in ref.\cite{Ratz}.

Let us now give an example which realizes our assumptions. One needs 
a $Z_2$-invariant neutrino mass matrix and a charged lepton mass
matrix which break it at the high scale. This breaking is required to 
be spontaneous. This can be done without invoking additional Higgs
doublets provided one introduces several singlet fields. The model below
is based on the MSSM augmented with two pairs of the  standard model
singlet fields denoted by $(\eta_1,\eta_2)$ and 
$(\bar{\eta}_1,\bar{\eta}_2)$. We impose a
discrete $Z_4\times Z_4$ symmetry under which various superfields
transform as follows:
\begin{eqnarray}
 \matrix{ (\mu^c,\tau^c,\bar{\eta}_1,\eta_1^*)\sim(i,1),&
(D_e,\eta_2,\bar{\eta}_2^*)\sim (1,i),& e^c\sim (1,-1),&D_ - \sim (-1,1) \cr
}~,
\label{symmetry}
\end{eqnarray}
where $D_\a (\a^c)$ denote the leptonic doublets
(singlets) with flavour $\alpha=e,\mu,\tau$; $D_\pm\equiv {D_\tau\pm
D_\mu\over \sqrt{2}}$. The $D_+$ and the standard Higgs superfields
$H_{u,d}$ transform as singlets.
We assume that the 
$Z_4\times Z_4$ symmetry is broken by the vacuum expectation values of the
$\eta$-fields at a scale only slightly
lower than
the neutrino mass scale $M_X$. As a result, non-renormalizable terms
involving these fields can
give sizable contributions to Yukawa couplings as in the Froggatt-Nielsen
mechanism \cite{FN}. 

The following dimension 5 terms in the superpotential contribute to the
charged lepton masses:
\begin{eqnarray}
\label{wl}
W_Y&=&D_+(\Gamma_\mu \mu^c +\Gamma_\tau \tau^c){H_d \eta_1\over
M_X}+D_ -(\Gamma_\mu' \mu^c +\Gamma_\tau' \tau^c){H_d
\bar{\eta}_1\over M_X}+\Gamma_e D_e e^c {H_d \eta_2 \over M_X}
~. \end{eqnarray}

The neutrino masses follow from the following non-renormalizable operators
invariant under the $Z_4\times Z_4$ symmetry: 
\begin{equation}
\label{wnu}
W_\nu={\alpha\over M_X} (D_+H_u)^T (D_+H_u)+{\beta \over M_X} (D_-H_u)^T
(D_-H_u)+{\gamma\over M_X} (D_+H_u)^T(D_e H_u) {\bar{\eta}_2\over M_X}~,
\end{equation}
where we have suppressed the Lorentz and $SU(2)$ indices.
Equation~(\ref{wl}) leads to the charged lepton mass matrix
\begin{eqnarray}
\label{chy}
{\mathcal M}_l &=&\left( \matrix{a_e&0&0\cr 0&a_\mu-a_\mu'&a_\tau-a_\tau'
\cr 0&a_\mu+a_\mu'&a_\tau+a_\tau'\cr}\right) ~,
\label{ml}
\end{eqnarray}
where 
\be
a_e=\Gamma_e\frac{\langle H_d^0\rangle \langle \eta_2\rangle}{M_X}~~~,~~~ 
a_\alpha={\Gamma_\a\over \sqrt{2}}{\langle
H_d^0\rangle\langle\eta_1\rangle\over M_X}~~~,~~~ 
a'_\alpha={\Gamma'_\a\over \sqrt{2}}{\langle
H_d^0\rangle\langle\bar{\eta}_1\rangle\over M_X} \ \ (\alpha=\mu,\tau)
~.
\ee
The neutrino mass matrix has the $Z_2$ invariant form of
equation~(\ref{mnu0f}) but with $X=0$. 
This together with the charged lepton mass matrix in equation~(\ref{chy})
imply that the $U_{e3}=0$ at the tree level. In the limit  
$a_\a=a_\a'$, equation~(\ref{chy}) leads to a massless muon and also corrections
to $\theta_{23}$ from the charged leptons vanish. In this limit, the model
is equivalent to the $Z_2$ model with $\gamma=\pi/4$. The
imposition of equality $a_\a=a_\a'$ is technically natural in the context
of supersymmetric theory. Small departure from it would lead to
the muon mass and a contribution  $\theta_{23
l}\approx \ord({m_\mu\over m_\tau})$ from the diagonalization of the
charged lepton matrix to $\theta_{23}$. In
this
case one gets the more general model represented by equation~(\ref{f1}). $U_{e3}$
still remains zero at $M_X$. 

The model discussed above reduces to the MSSM below the $Z_4\times Z_4$
breaking scale. $\mnu$ in this case is
invariant under a $Z_2$ symmetry which interchanges $D_\mu$ with $D_\tau$.
This $Z_2$ however is not a symmetry of the charged lepton Yukawa
couplings, equation (\ref{wl}). Even in the neutrino sector,  
the $Z_2$ invariance is only approximate one and is broken by the terms
of $\ord(\frac{\langle\eta\rangle ^2}{M_X^2})$ where $\langle\eta\rangle$
generically denotes the
vacuum expectation value for any of the singlet fields. The parameter
$\lambda\sim
\frac{\langle\eta\rangle}{M_X}$ determines the tau lepton mass in equation 
(\ref{wl}) and is
required to be $\geq\ord(10^{-2})$ if the Yukawa couplings
$\Gamma_{\alpha}$ are to remain below 1. This means that the neglected
non-leading terms
in 
equations (\ref{wl},\ref{wnu}) are typically  $\ord( 10^{-2})$ smaller
than the leading ones. 

The breakdown of the $Z_2$ symmetry and a non-zero $U_{e3}$ arise in the
model from the non-leading terms not displayed in
equations (\ref{wl},\ref{wnu}). The charged lepton mass matrix gets
additional
contributions from the following $Z_4\times Z_4$ invariant dimension six
terms in the super potential:
\begin{equation}
\label{wfnl}
 D_e(\beta_{e\mu} \mu^c+\beta_{e\tau}\tau^c)H_d
\frac{\eta_1\bar{{\eta}}_2}{M_X^2}+D_+e^cH_d\frac{\beta_e\eta_2^2+
\bar{\beta}_e\bar{\eta}_2^2}{M_X^2}
~.\end{equation}
The corrected charged lepton mass matrix then has the following form
\begin{eqnarray}
\label{ml6}
 {\mathcal M}_l= \left (\matrix{
a_e&\lambda_{e\mu}&\lambda_{e\tau}\cr
\lambda_e&a_\mu-a_\mu'&a_\tau-a_\tau'\cr
\lambda_e&a_\mu+a_\mu'&a_\tau+a_\tau'\cr}\right )~.
\end{eqnarray}
Here, $\lambda_{e,e\mu,e\tau}$ can be read-off from equation (\ref{wfnl}).
These are suppressed compared to the leading terms in equation (\ref{wl})
by $\lambda=\frac{\langle \eta \rangle}{M_X}$ where $\langle \eta\rangle$
refers to a typical vacuum expectation of any of the singlet fields.
An estimate of $\lambda$ can be obtained by noting that it
determines the tau lepton mass in equation 
(\ref{wl}) and is
required to be $\geq\ord(10^{-2})$ if the Yukawa couplings
$\Gamma_{\alpha}$ are to remain below 1. This means that 
the terms $\lambda_{e\alpha},\lambda_e$ in equation (\ref{ml6})
can be $\ord(m_\mu)$ if the relevant Yukawa couplings
are
$\ord(1)$. They can therefore significantly affect the $e-\mu$
sector and would lead to a large electron mass and $e-\mu$ mixing.
This requires assuming suppression in some of the Yukawa couplings.
While different choices are possible, we give an example which is
particularly interesting. This corresponds to choosing
$a_e\ll m_e; a_\alpha=a_\alpha'\approx
\ord(m_\tau)~;\lambda_{e\tau}\sim\lambda_{e\mu}\sim \ord
(m_e)$
and $\lambda_e\sim \ord(m_\mu)$. The $\lambda_{e \alpha}$ contribute
to the electron mass and the corresponding Yukawa couplings
$\beta_{e\alpha}$ need to be suppressed $\beta_{e\alpha}\sim
\ord(\frac{m_e}{m_\mu})$. One gets correct pattern for the charged
lepton masses and a contribution of $\ord(\frac{m_e}{m_\tau})$ to $U_{e3}$
from the charged lepton sector. 
The radiatively
induced $U_{e3}$ can be larger than this as seen from Fig. 12.

The non-leading terms break $Z_2$  in the neutrino sector also
and lead to a direct contribution to $U_{e3}$. 
This comes from the terms of the type
\begin{equation}
\frac{(\eta_1^2,\bar{\eta}_1^2)}{M_X^3} (D_+H_u)^T D_-H_u~~;~~ 
\frac{\bar{\eta}_2}{M_X^4}(\eta_1^2,\bar{\eta}_1^2) (D_-H_u)^T
D_eH_u~~;~~ 
\frac{(\eta_2^2,\bar{\eta}_2^2)}{M_X^3} (D_eH_u)^T D_eH_u~.
\end{equation}
These terms are typically suppressed by $\ord(10^{-2})$ compared to the
corresponding leading terms displayed in equation (\ref{wnu}).

\section{Conclusions}

The neutrino mixing matrix contains two small parameters $\ue3$ and $\cos
2\theta_{23}$ which would influence the outcome of the future neutrino
experiments. This paper was devoted to study of these parameters within a
specific theoretical framework. The vanishing of $\ue3$ was shown to
follow from a class of $Z_2$ symmetries of ${\mathcal M}_{\nu f}$. 
This symmetry
can be used to parameterize all models with zero $U_{e3}$. A specific
$Z_2$ in this class also leads to the maximal atmospheric neutrino mixing
angle. We showed that breaking of this can be characterized by two
dimensionless parameters $\e,\e'$ and we studied their effects
perturbatively and numerically.

It was found that the magnitudes of $\ue3$ and  $|\cos 2\theta_{23}|$
are  strongly dependent upon the neutrino mass hierarchies
and CP violating phases. The $\ue3$ gets strongly suppressed in case of 
the inverted or quasi degenerate neutrino spectrum if $\rho=\sigma$ while
similar suppression occurs in the case of normal hierarchy independent of 
this phase choice. The choice $\rho\not =\sigma$ can lead to a larger
values $\sim 0.1$ for $\ue3$ which could be close to the experimental
value in some cases with inverted or quasi-degenerate spectrum.
 In contrast, the $|\cos 2\theta_{23}|$ could be
large, near its present experimental limit in most cases studied. For the
normal and inverted mass spectrum, the magnitude of  $\cos 2\theta_{23}$
is similar to the magnitudes of the perturbations $\e,\e'$ while it can
get enhanced compared to them if the neutrino spectrum is
quasi-degenerate.

The phenomenological implications of the present scheme are distinct from
various other schemes discussed in the literature
\cite{bd,nubim,lbim,sm,as,gou}.
Ref. \cite{gou} considered various neutrino mass textures which lead to
zero solar scale, $U_{e3}=0$ and $\cos 2\theta_{23}=0$,  and applied random
perturbations to them. In this approach, both $\ue3$ and $|\cos
2\theta_{23}|$ were found to be similar in contrast to the present case
which predicts $\ue3\leq |\cos 2\theta_{23}|$. The approach of \cite{gou} 
predicts 
large  $|\cos2\theta_{23}|$ of $\ord(\sqrt{\soa})$ for the normal neutrino
mass
hierarchy and small $\ord(\soa)$ in the other cases. This is quite
different from our results as seen in Table 1.

An alternative proposal is to make assumptions on the leptonic mixing
matrices $U_{\nu,l}$. The cases considered correspond to a bi-maximal form
for $U_\nu$ with a small corrections from $U_l$ \cite{nubim} or its
converse \cite{lbim}. If $U_\nu$ is bi-maximal and $U_l$ gives small
corrections than one finds rather large $\ue3$ near the present limit and
moderate $|\cos 2\theta_{23}|$, {\it e.g.}, $|\cos 2 \theta_{23}|\leq 0.12$
in the specific scheme considered in \cite{sm}. The converse case with
the bi-maximal $U_l$ and $U_\nu$ with a typical form of the CKM matrix
is characterized by small $\ue3\sim 0.02$ and small
 $|\cos 2\theta_{23}|\leq 0.08$ \cite{sm}.   

One sees clear distinctions in the predictions of
various models and it should be possible to rule out some of them
once the challenging task of the experimental  determination of $\ue3$ and
$|\cos 2\theta_{23}|$ is  accomplished.
\vskip 0.5 cm
 \textbf{Note}:
After this work was completed, we found a paper by Mohapatra with the similar
 discussion based on the $\mu$--$\tau$ interchange symmetry \cite{Moh} .

\vskip 0.2 cm
  \textbf{Acknowledgments}:
The work of L.L.\ has been supported
by the Portuguese \textit{Funda\c c\~ao para a Ci\^encia e a Tecnologia}
under the project CFIF--Plurianual.
M.T.\ has been supported
by the Grant-in-Aid for Science Research
from the Japanese Ministry of Education, Science and Culture
No.\ 12047220, 16028205. S.K. is also supported by 
the Japan Society for the Promotion of Science (JSPS).
 A.S.J. would like to thank  the JSPS  for a grant which made
this collaboration possible. 

\newpage

\end{document}